\documentclass[aps,twocolumn,showpacs,preprintnumbers,floatfix]{revtex4}
\usepackage{graphicx}
\usepackage{epsfig}
\usepackage{dcolumn}
\usepackage{subfigure}
\usepackage{amsmath}

\def\be{\begin{equation}}
\def\ee{\end{equation}}
\def\bea{\begin{eqnarray}}
\def\eea{\end{eqnarray}}

\def\d#1#2{\frac{\displaystyle #1}{\displaystyle #2}}
\def\no{\nonumber}
\def\p{\partial}
\newcommand{\omits}[1]{}

\def\bsp{\be\begin{split}}

\def\bes{\be  \begin{split}}

\def\p{\partial}

\newcommand{\Rmnum}[1]{\expandafter\@slowromancap\romannumeral #1@}

\def\PRD{{Phys. Rev.}~{\bf D}}
\def\PRL{{Phys. Rev. Lett. }}

\def\CQG{{Class. Quant. Grav. }}

\def\JHEP{{JHEP}}

\begin{document}

\title{Stability of black holes based on horizon thermodynamics}
\author{Meng-Sen Ma$^{a,b}$\footnote{Email: mengsenma@gmail.com; ms\_ma@sxdtdx.edu.cn}, Ren Zhao$^{a,b}$}

\medskip

\affiliation{\footnotesize$^a$Department of Physics, Shanxi Datong
University,  Datong 037009, China\\
\footnotesize$^b$Institute of Theoretical Physics, Shanxi Datong
University, Datong 037009, China}

\begin{abstract}
On the basis of horizon thermodynamics we study the thermodynamic stability of black holes constructed in general relativity and Gauss-Bonnet gravity.
In the framework of
horizon thermodynamics there are only five thermodynamic variables $E,~P,~V,~T,~S$. It is not necessary to consider concrete matter fields, which may contribute to the pressure of black hole thermodynamic system.  In non-vacuum cases, we can derive the equation of state, $P=P(V,T)$. According to the requirements of stable equilibrium in conventional thermodynamics, we start from these thermodynamic variables to calculate the heat capacity at constant pressure and Gibbs free energy and analyze the local and global thermodynamic stability of black holes. It is shown that $P>0$ is the necessary condition for black holes in general relativity to be thermodynamically stable, however this condition cannot be satisfied by many black holes in general relativity. For black hole in Gauss-Bonnet gravity negative pressure can be feasible, but only local stable black hole exists in this case.

\end{abstract}

 \maketitle

\section{Introduction}

Since the discovery of Hawking radiation we know that black holes have temperature.
Thus, the concept of entropy for black holes proposed by Bekenstein is no longer an anologue.
Not only that, the works of Hawking made the entropy quantitative, namely $S=A/4$. The laws of
black hole mechanics\cite{Hawking:1973} plus the generalized second law of thermodynamics(GSL)\cite{Bekenstein:1974}
imply that black holes are thermodynamic systems. The work of Jacobson\cite{Jacobson:1995} by deriving the Einstein equation from a thermodynamic equation of state
and the work of Padmanabhan\cite{Pa1,Pa2} by writing Einstein¡¯s equations for a spherically symmetric
spacetime in the form of the first law of thermodynamics make the connection between gravity and thermodynamics very close.
However, there are some differences between black hole thermodynamics and conventional thermodynamics, such as the black hole entropy, which
is proportional to the horizon area but not volume, and the heat capacity of black holes which may be negative.

Based on the general first law of black hole thermodynamics: $dM=TdS+\Omega dJ+\cdots$,
where ``$\cdots$" denote the possible additional contributions from long range fields, many thermodynamic properties of black holes have been studied.
The couplings of different matter fields to gravity will give different black hole solutions, such as in general relativity (GR) Maxwell field leads to RN black hole and Born-Infeld (BI) field gives the BI black hole. And these black holes derived in the same gravitational theory with different matter fields have very different dynamical and thermodynamical properties.

In fact, we can also treat the black hole thermodynamic system from another perspective. According to the horizon thermodynamics proposed by Padmanabhan,  one can obtain the thermodynamic identity: $dE=TdS-PdV$ from the field equations. One should also analyze the thermodynamic properties of black holes on the basis of this identity. In horizon thermodynamics, the pressure $P$ is the $(rr)$ component of energy-momentum tensor and the concrete form of it is unnecessary. In other words, the thermodynamic properties is relevant to the gravitational theories under consideration but insensitive to the concrete black hole solutions. In this paper, we will analyze the thermodynamic stability of black holes in the framework of horizon  thermodynamics. In this case, only two pairs of thermodynamic variables exist, which are the intensive quantities $T,~P$ and the extensive quantities $S,~V$. One can expect to obtain some different results from those obtained in usual black hole thermodynamics.

The paper is arranged as follows. In Section II, we simply introduce the horizon thermodynamics and black hole thermodynamics and demonstrate that they can be derived from the same field equation. In Section III we will
analyze the thermodynamic stability of black holes in GR. The thermodynamic properties of Gauss-Bonnet black hole is discussed in Section IV. We  make some
concluding remarks in Section V.

\section{horizon thermodynamics and black hole thermodynamics from Einstein field equation}

In this section, we take the static spherically symmetric black hole in general relativity as an example
and demonstrate that two forms of black hole thermodynamics can be derived from the same field equations.

For a static, spherically symmetric spacetime, the metric in the Schwarzschild gauge can be written as
\be\label{staticmetric}
ds^2=-f(r)dt^2+f(r)^{-1}dr^2 + r^2d\Omega^2.
\ee
For simplicity, one can consider another form of the metric function:
\be\label{fmr}
f(r)=1-{2m(r)\over r},
\ee
where $m(r)$ is an effective mass function. Obviously, for black holes with the metric taking the form of  Eq.(\ref{fmr}), the event
horizon lies at $r_{+}=2m(r_{+})$.

Substituting it into the Einstein field equation
\be
G^\mu_{~\nu}=R^\mu_{~\nu}-\d{1}{2}R g^\mu_{~\nu}=8\pi T^\mu_{~\nu},
\ee
the ``00" and ``11" components are the same and are expressed as
\be\label{mr}
\d{d m(r)}{d r}=-4\pi r^2 T^0_{~0}.
\ee
One can easily find that the ``22" and ``33" components of Einstein equations are satisfied by Eq.(\ref{mr}) automatically. It should be noted that although we do not explicitly introduce the cosmological constant $\Lambda$ in the theory, it can be included in the $T_{\mu\nu}$.

The Hawking temperature of black holes can be calculated according to
\bea\label{tem}
T&=&\d{\kappa}{2\pi}=-\left.\d{1}{4\pi}\d{\p_{r}g_{tt}}{\sqrt{-g_{tt}g_{rr}}}\right|_{r=r_{+}}\no \\
&=&\left.\d{1}{4\pi}f'(r)\right|_{r=r_{+}}=\d{1}{4\pi r_{+}}-\d{m'(r_{+})}{2\pi r_{+}}.
\eea
Thus, Eq.(\ref{mr}) can be rewritten as
\be\label{ht}
\d{1}{2}=2\pi r_{+}T-4\pi r_{+}^2T^0_{~0}.
\ee
Multiplying $dr_{+}$ on both sides of the above equation and integrating them, we can obtain
\be\label{ht-bh}
dE=TdS-PdV,
\ee
where $E=r_{+}/2$ is the internal energy of the system; $V=4\pi r_{+}^3/3$ is the thermodynamic volume of the black hole; $P=T^0_{~0}=T^1_{~1}$ is the pressure; and
$S=\pi r_{+}^2$ is the entropy of the black hole. Thus, we obtain the thermodynamic identity from gravitational field equation. This is the horizon thermodynamics of black holes in GR.

Next, we continue to derive the first law of black hole thermodynamics.
Integrating Eq.(\ref{mr}), one can obtain
\be\label{mr2}
m(r)=M+4\pi\int_{r}^{\infty}r^2T^0_{~0}dr,
\ee
where the integration constant is chosen according to the requirement $M=\lim\limits_{r\rightarrow \infty}m(r)$.
At the horizon, one can derive from Eq.(\ref{mr2})
\be
M=\d{r_{+}}{2}-4\pi\int_{r_{+}}^{\infty}r^2T^0_{~0}dr
\ee
Taking the variation of the above equation, one can obtain
\be
\delta M=\d{1}{2}\delta r_{+}-4\pi \delta \int_{r_{+}}^{\infty}r^2T^0_{~0}dr.
\ee
It should be noted that if $T^0_{~0}=T^0_{~0}(r,Q)$, namely $T^0_{~0}$ is not only a function of $r$ but also contains charges of some matter fields, one can obtain\cite{Ma}
\be
\delta M=\left(\d{1}{4\pi r_+}+2r_{+}T_{~0}^{0}\right)\delta \d{A}{4}-\left(4\pi\int_{r_{+}}^{\infty}r^2\d{\p T^0_{~0}}{\p Q}dr\right)\delta Q.
 \ee
This is just the usual first law of black hole thermodynamics,
\be\label{1st-bh}
\delta M=T\delta S+\phi \delta Q
 \ee
 with $\phi=-4\pi\int_{r_{+}}^{\infty}r^2\d{\p T^0_{~0}}{\p Q}dr$ is the conjugate potential of the charge $Q$. For example, $T^\mu_{~\nu}=\d{Q^2}{8\pi r^4}diag(-1,-1,1,1)$ for RN black hole. Thus, we can easily calculate
\be
\phi=-4\pi\int_{r_{+}}^{\infty}r^2\d{\p T^0_{~0}}{\p Q}dr=\d{Q}{r_{+}},
\ee
which is just the electric potential measured at infinity with respect
to the horizon.

The truth that Eq.(\ref{ht-bh}) in the horizon thermodynamics and Eq.(\ref{1st-bh}) in the usual black hole thermodynamics can be derived from the same field equation indicates that the two formulae can be derived each other. Based on the usual black hole thermodynamics, the thermodynamic properties including the thermodynamic stability of many black holes have been studied extensively. However, we will show below that the thermodynamic properties of black holes based on the horizon thermodynamics are very different.

\section{thermodynamic stability of black holes in GR}

One can see that Eq.(\ref{ht}) is in fact an equation of state described by three state parameters $P,~V,~T$, which is
\bea\label{GRPVT}
P&=&\d{T}{2 r_{+}}-\d{1}{8\pi r_{+}^2}\no \\
 &=&\d{(4\pi)^{1/3}T}{2\times3^{1/3}V^{1/3}}-\d{1}{2\times3^{2/3}\times(4\pi)^{1/3}V^{2/3}}.
\eea
Stable equilibrium for a thermodynamic system requires that $\left.\d{\p P}{\p V}\right|_{T}\leq 0$ or  $\kappa_T\geq 0$\cite{Callen}. We can calculate it easily
\be\label{pvt}
\left.\d{\p P}{\p V}\right|_{T}=\frac{\sqrt[3]{6}-(6 \pi )^{2/3} T \sqrt[3]{V}}{18 \sqrt[3]{\pi } V^{5/3}}=\d{\sqrt[3]{6} \left(1-2 \pi  r_+ T\right)}{18 \sqrt[3]{\pi } V^{5/3}}.
\ee
Thus, to guarantee the stability of the thermodynamic system there must be $T \geq\d{1}{2\pi r_{+}}$.
From Eq.(\ref{ht}), we know that $T =2 P r_++\frac{1}{4 \pi  r_+}$. This requirement means that $P\geq \d{1}{8\pi r_{+}^2}>0$. Here we should stress that the discussion above is based on the non-vacuum case, namely nonzero $P$. Obviously, in the vacuum case, such as Schwarzschild solution, the thermodynamic quantities $V,~T,~S$ are not independent and thus the LHS of Eq.(\ref{pvt}) cannot be defined.

Another requirement for stable equilibrium is $C_P\geq C_V\geq 0$\cite{Callen}. According to general definition,
\be
C_{V}=\left.\d{\p E}{\p T}\right|_{V}=\left.T\d{\p S}{\p T}\right|_{V}=0,
\ee
because constant $V$ means constant $E$ and $S$ for black hole thermodynamic system.
We can only define the heat capacity at constant pressure, which is
\be
C_{P}=\left.\d{\p H}{\p T}\right|_{P}=\left.T\d{\p S}{\p T}\right|_{P}=\frac{2 \pi  r_+^2 \left(8 \pi  P r_+^2+1\right)}{8 \pi  P r_+^2-1},
\ee
where $H=E+PV$ is the enthalpy of the system.

\begin{figure}[!htbp]
\includegraphics[width=7cm,keepaspectratio]{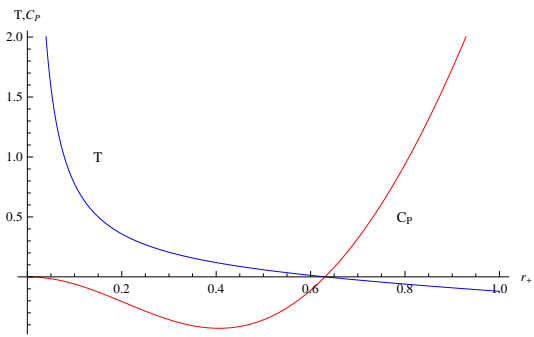}\hspace{0.5cm}
\caption{The heat capacity and temperature as functions  of  $r_{+}$ for negative pressure.\label{TCP}
}
\end{figure}

Clearly, $C_{P}$ is always negative when $P=0$. As is shown in Fig.\ref{TCP},  $C_{P}$ can be positive for negative $P$ when the horizon radius is larger than a critical value. However, one can find that in this case the temperature is negative. Therefore, to guarantee $C_{P}\geq0$  and physically meaningful thermodynamic quantities, there must be $P>0$.
Therefore, black holes derived from general relativity coupled with the matter fields with $P\leq 0$ are thermodynamically unstable.

\begin{figure}[!htbp]
\includegraphics[width=7cm,keepaspectratio]{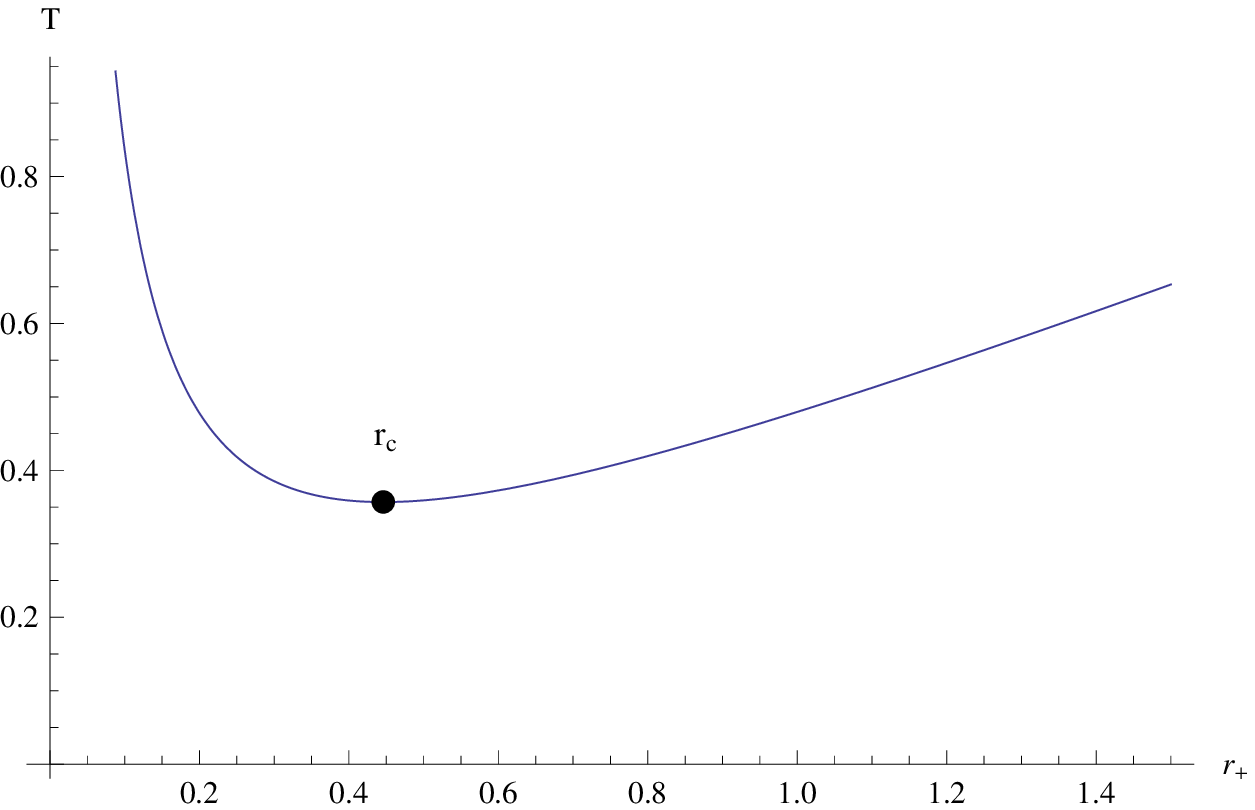}\hspace{0.8cm}\\
\includegraphics[width=7cm,keepaspectratio]{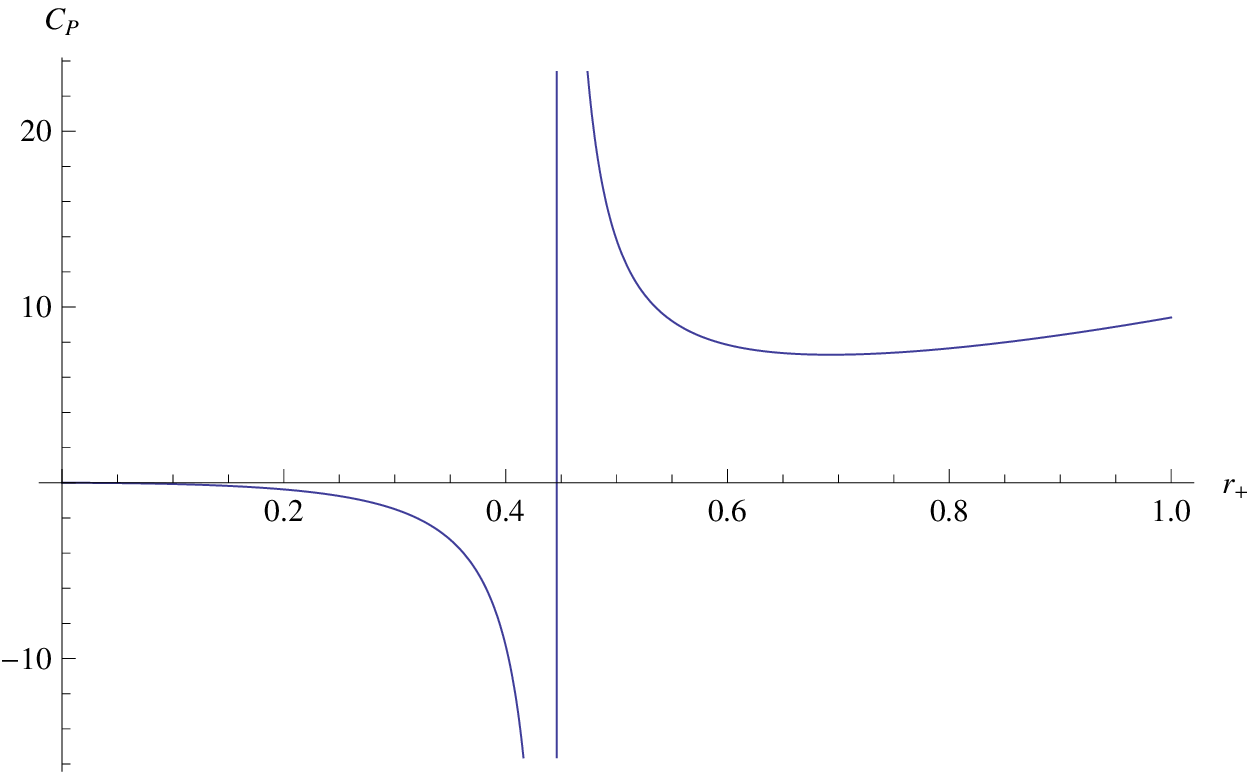}
\caption{The Hawing temperature and the heat capacity at constant pressure as function  of  $r_{+}$ for $P=0.2$. At $r_{+}=r_c=0.446$, the temperature takes minimum and correspondingly the heat capacity diverges.\label{pp}
}
\end{figure}

Below we only consider the case with $P>0$. In this case, the temperature has the minimum $T_{min}=\sqrt{2P/\pi}$. As is depicted in Fig.\ref{pp}, the temperature is always positive and for the given value $P=0.2$ we can find that the heat capacity $C_P$ diverges at $r_{+}=r_c=0.446$ where the temperature takes minimum. According to the viewpoint of Davies\cite{Davies}, the divergence of heat capacity means the second-order phase transition happens there. The heat capacity is positive for larger black hole with $r_{+}>r_c$ and negative for the smaller black hole with $r_{+}<r_c$. It means that the larger black hole is thermodynamically stable locally.

Similarly, one can also calculate the compressibility $\kappa_T$ and the expansion coefficient $\alpha$:
\bea
\kappa_T&=&-\d{1}{V}\left.\d{\p V}{\p P}\right|_{T}=\frac{24 \pi  r_+^2}{8 \pi  P r_+^2-1}, \\
\alpha &=&\d{1}{V}\left.\d{\p V}{\p T}\right|_{P}=\frac{12 \pi  r_+}{8 \pi  P r_+^2-1}.
\eea
Obviously, the two quantities have the same divergent point as the heat capacity at constant pressure.

One can also calculate the Gibbs free energy to discuss the global stability of the black holes in GR.  We can define it as
\be
G=E-TS+PV = \frac{r_+}{4}-\frac{2}{3} \pi  P r_+^3.
\ee

\begin{figure}[!htbp]
\subfigure[$G-r_{+}$] {
\includegraphics[width=7cm,keepaspectratio]{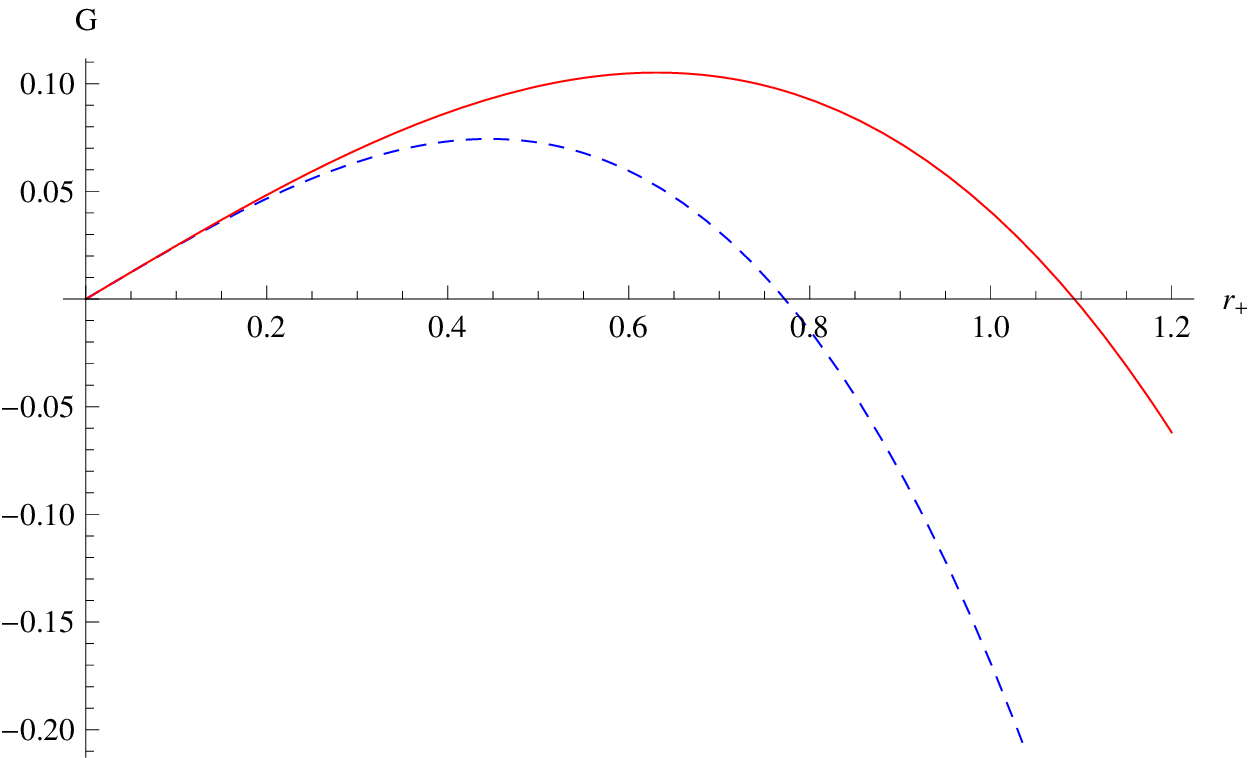}\label{G1}}
\subfigure[$G-T$] {
\includegraphics[width=7cm,keepaspectratio]{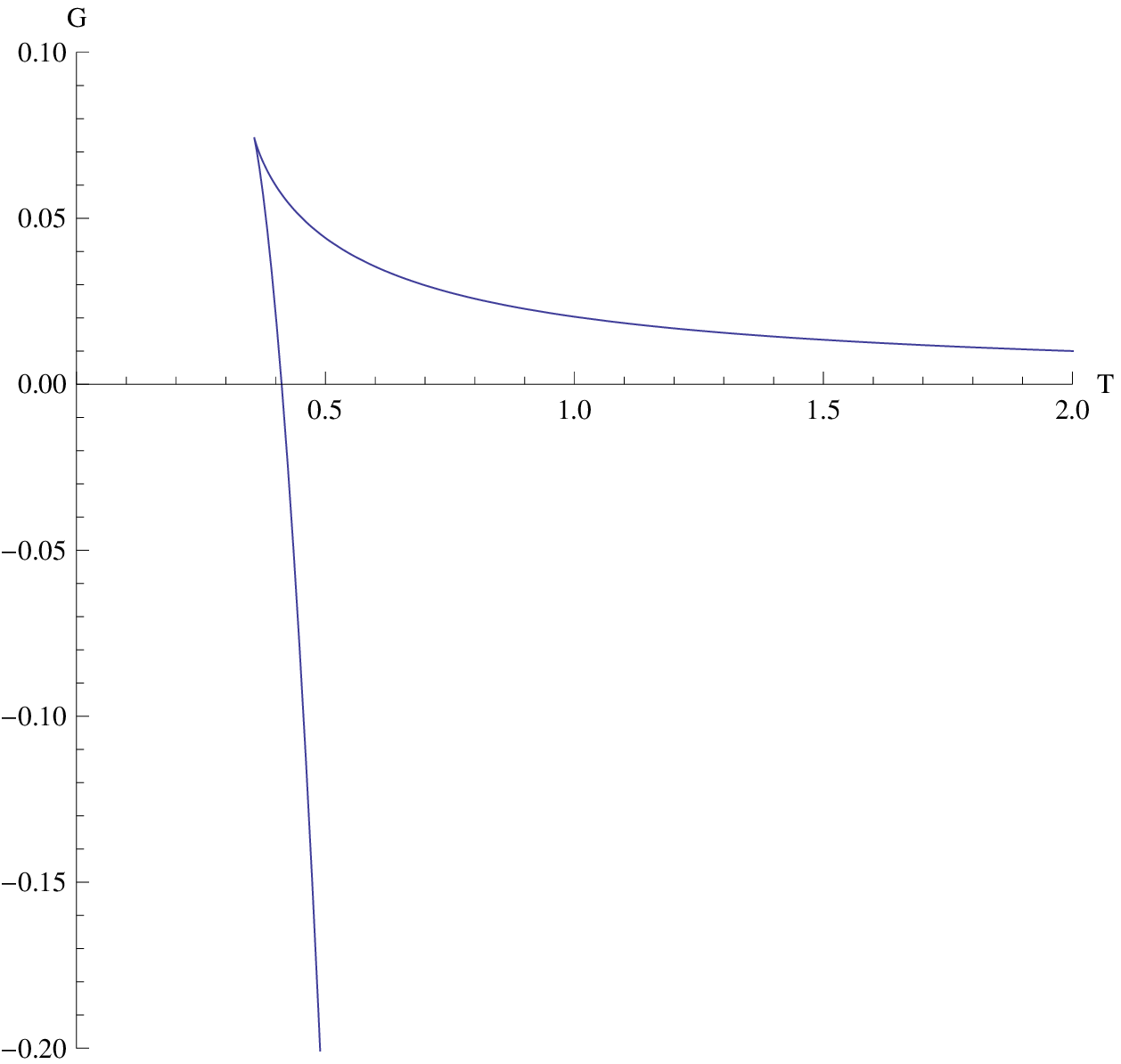}\label{G2}}
\caption{In (a), the dashed (blue) line corresponds to the case $P=0.2$ and the solid line (red) corresponds to $P=0.1$. The $G-T$ figure is depicted with $P=0.2$.}
\end{figure}
As is shown in Fig.\ref{G1}, the Gibbs free energy has maximum at the divergent point of $C_{P}$. On the right hand side of that point, $G$ is a monotonically decreasing function of $r_{+}$ for constant pressure $P$. Thus, we know that,
the larger the black hole is, the more stable it will be when the horizon radius is larger than the critical value $r_c$. While for fixed $r_{+}$, the black holes with larger pressure is more stable. Fig.\ref{G2} shows that for black holes in GR with nonzero matter sources there is a Hawking-Page-like phase transition in the framework of horizon thermodynamics, which is a kind of radiation/large black hole phase transition\cite{Hawking-Page}. For more details, one can consult the work\cite{Kubiznak}, where all the thermodynamic quantities for Schwarzschild-AdS black hole are given. It can be found that the thermodynamic quantities obtained for black holes in GR in the framework of horizon thermodynamics are indeed the same as those given in \cite{Kubiznak}.

The above results apply for all black holes in GR. Now we see some specific black holes as examples.

(1)RN black hole. It is a solution of GR with electromagnetic field as source. The energy density $\rho=-T^0_{~0}=\d{Q^2}{8\pi r^4}>0$ and the radial pressure $P=T^{1}_{~1}=\d{-Q^2}{8\pi r^4}<0$. According to discussion above, RN black hole cannot be thermodynamically stable in the framework of horizon thermodynamics. This result is very different from those obtained in the usual black hole thermodynamics\cite{Pavon,Chamblin,Lemos,Wu1,Lundgren,Banerjee,Wu2}. To guarantee thermodynamic stability, there must be other special matter contained in the sources. The simplest choice is the cosmological constant. In fact, RN-AdS black hole can be in stable equilibrium because $P=-\d{Q^2}{8\pi r^4}-\d{\Lambda}{8\pi}$ may be positive in this case.

(2)Born-Infeld black hole. The Lagrangian is given by $L(F)=4\beta^2\left(1-\sqrt{1+F_{\mu\nu}F^{\mu\nu}/2\beta^2}\right)$, where the Born-Infeld parameter $\beta$ has the dimension of mass. In the limit $\beta \rightarrow \infty$, Born-Infeld theory reduces to the usual Maxwell electromagnetic theory. One can derive the energy-momentum tensor and obtain the radial pressure $8\pi P=2\beta^2\mp2\beta\sqrt{\beta^2r^4+Q^2}/r^2$, where ``$-$" for positive $\beta$ and ``$+$" for negative $\beta$. Therefore the radial pressure is also negative for BI black hole and thus  BI black hole is also thermodynamically unstable in the sense of horizon thermodynamics.

In fact, from our assumption $g_{00}=-g_{11}^{-1}$ in the metric form, there must be $T^0_{~0}=T^1_{~1}$ in GR. According to energy conditions, there is always $\rho=-T^0_{~0}\geq 0$, thus
$P=p_r=T^1_{~1}\leq0$. Except for the vacuum case $P=0$, we find that all static spherically symmetric black holes in GR may be thermodynamically unstable in the framework of horizon thermodynamics. Next we will consider modified theory of gravity to further study the stability of black holes.

\section{thermodynamic stability and critical behavior of Gauss-Bonnet black hole}

For Gauss-Bonnet(GB) gravity, the action is given by
\be
\mathcal{S}=\d{1}{16\pi}\int d^5x\sqrt{-g}\left[R+\alpha\mathcal{L}_{GB}\right]+\mathcal{S}_{matter}.
\ee
We have set the dimension of spacetime $D=5$. $\alpha$ is the Gauss-Bonnet coefficient and is positive according to string theory\cite{Deser:1985}.$\mathcal{L}_{GB}$ is the Gauss-Bonnet term and takes the form
\be
\mathcal{L}_{GB}=R^2-4R_{\mu\nu}R^{\mu\nu}+R_{\mu\nu\gamma\delta}R^{\mu\nu\gamma\delta}.
\ee
Considering again a static, spherically symmetric metric
\be\label{staticmetric5}
ds^2=-f(r)dt^2+f(r)^{-1}dr^2 + r^2d\Omega_{3}^2,
\ee
and substituting this into field equations, one can obtain
\be\label{GB11}
f'(r_{+}) \left(\frac{4 \alpha }{r_+}+r_+\right)-2=\frac{16 \pi}{3} r_+^2 T^{r}_{~r}.
\ee
As is shown in \cite{Pad:2006}, this equation can also be rewritten into the form $dE=TdS-PdV$ with
\bea
E&=&\d{3\pi r_{+}^2}{8}\left(1+\d{2\alpha}{r_{+}^2}\right), \quad S= \d{\pi^2r_{+}^3}{2}\left(1+\d{12\alpha}{r_{+}^2}\right),\no \\
V&=&\d{\pi^2r_{+}^4}{2}, \quad T=\d{f'(r_{+})}{4\pi}
\eea

We can obtain the Hawking temperature from Eq.(\ref{GB11})
\be
T=\frac{r_+ \left(8 \pi  P r_+^2+3\right)}{6 \pi  \left(4 \alpha +r_+^2\right)}.
\ee
\begin{figure}[!htbp]
\includegraphics[width=7cm,keepaspectratio]{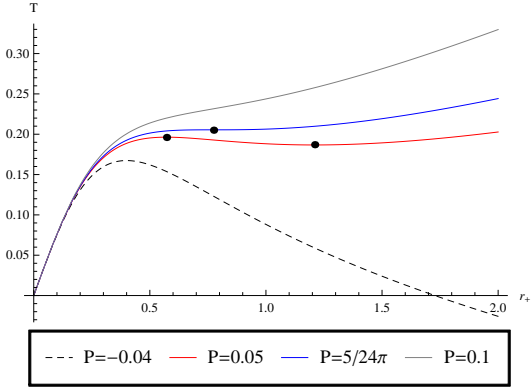}\hspace{0.5cm}
\caption{The Hawking temperature $T$ of Gauss-Bonnet black hole as functions  of  $r_{+}$ for different $P$. We have set $\alpha=0.05$.\label{Tgb}
}
\end{figure}
As is shown in Fig.\ref{Tgb},  the Hawking temperature with positive pressure is always positive. While the Hawking temperature with negative pressure is positive for smaller $r_{+}$ and negative for larger $r_{+}$. Although we think that negative temperature for black holes is meaningless, we cannot exclude the case with negative pressure due to the presence of positive temperature part. Moreover, we can find that when $P<\frac{5}{24 \pi }$ the temperature has a local minimum and a local maximum, when $P=\frac{5}{24 \pi }$ the two local extrema coincide.
When $P>\frac{5}{24 \pi }$ the temperature is a monotonically increasing function of $r_{+}$ for constant pressure $P$.

\begin{figure}[!htbp]
\includegraphics[width=7cm,keepaspectratio]{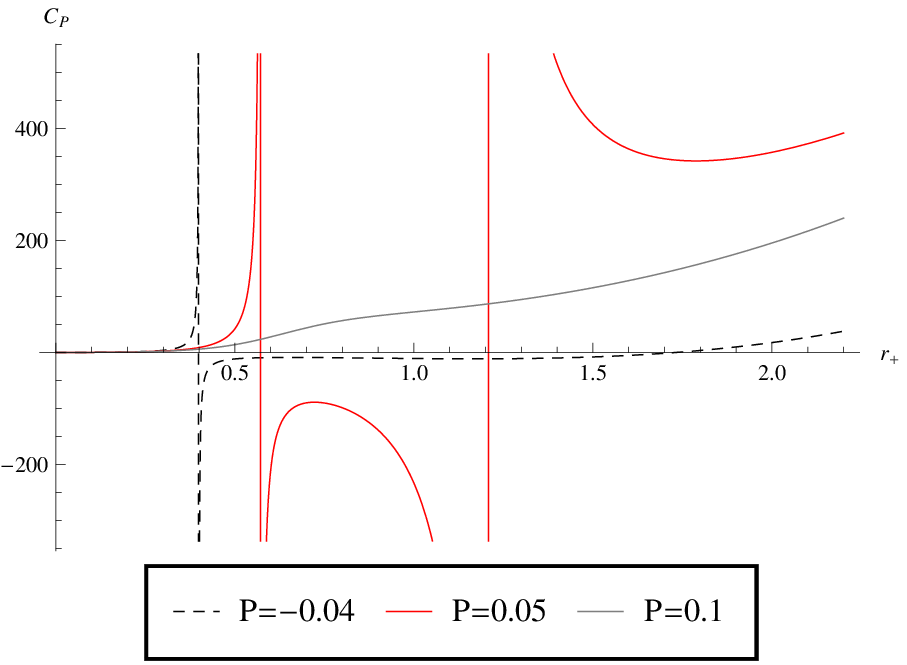}\hspace{0.5cm}
\includegraphics[width=7cm,keepaspectratio]{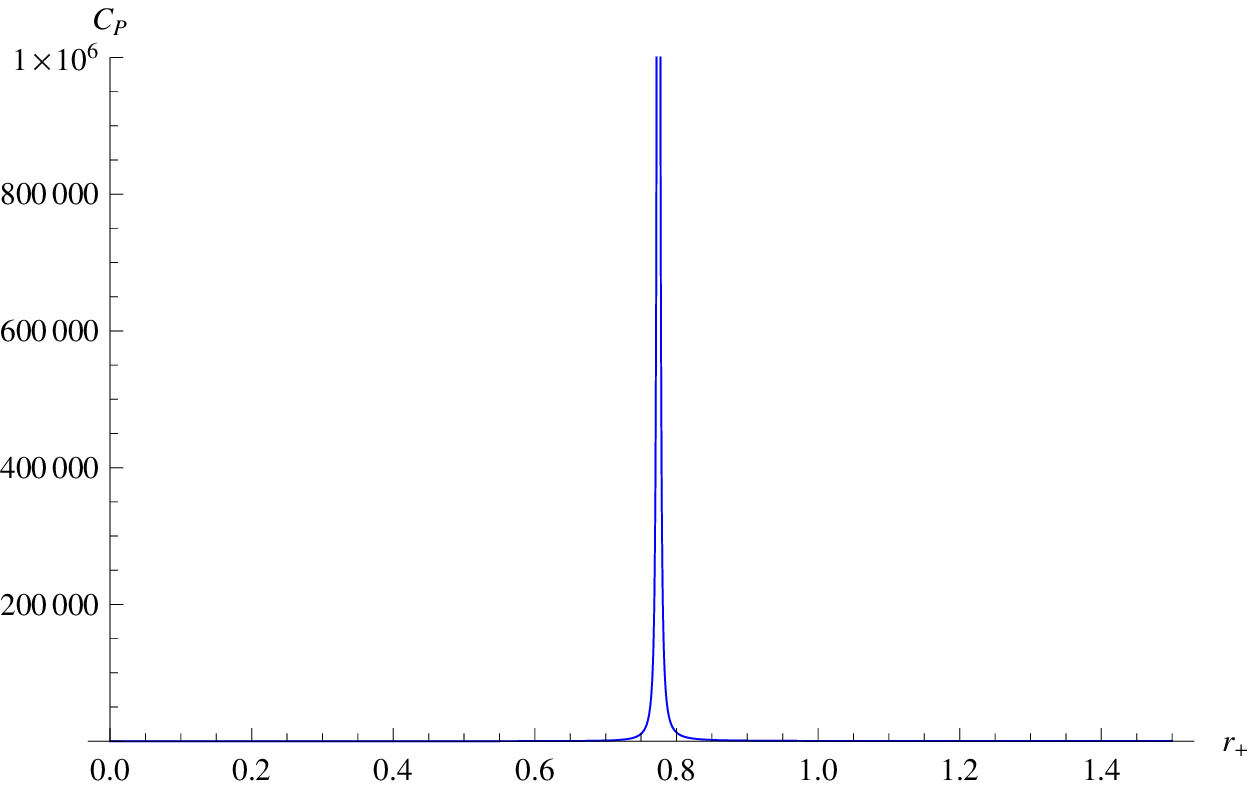}
\caption{The heat capacity at constant pressure $C_{P}$ of Gauss-Bonnet black hole as functions  of  $r_{+}$ for different $P$. The picture below corresponds to the case $P=5/24\pi$. We have set $\alpha=0.05$.\label{Cpgb}
}
\end{figure}
Similar reason to that in GR, $C_V$ is also zero for GB black hole. We can only derive the heat capacity at constant pressure:
\be
C_{P}=\frac{3 \pi ^2 r_+ \left(8 \pi  P r_+^2+3\right) \left(4 \alpha +r_+^2\right){}^2}{2 \left(12 \alpha +r_+^2 (96 \pi  \alpha  P-3)+8 \pi  P r_+^4\right)}.
\ee
Corresponding to the temperatures, there are two or one or no divergent points for different choices of $P$, which is depicted in Fig.\ref{Cpgb}.

When $P<0$, $C_P$ is positive for smaller black hole and negative for intermediate black hole, and is again positive for larger black hole. However, for the larger black hole the temperature is negative. Therefore, in this case, only the smaller black hole is locally stable.
When $0<P<\frac{5}{24 \pi }$, there are two divergent points for $C_P$, which correspond to the two extrema of temperature. Clearly, the smaller and larger black holes are both local stable due to the positivity of $C_P$, while the intermediate black hole is unstable.
When $P=\frac{5}{24 \pi }$, the two divergent points coincide. When $P>\frac{5}{24 \pi }$, $C_P$ is always positive and no divergent point exists. This means that in this case the Gauss-Bonnet black hole is local stable for any values of $r_{+}$.

\begin{figure}[!htbp]
\subfigure[$G-r_{+}$ for GB black hole] {
\includegraphics[width=7cm,keepaspectratio]{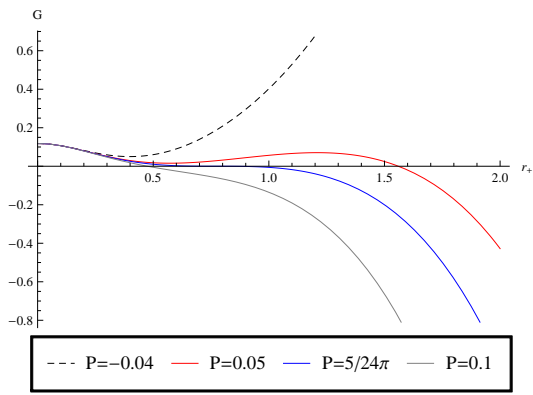}\label{G3}}
\subfigure[$G-T$ for GB black hole] {
\includegraphics[width=7cm,keepaspectratio]{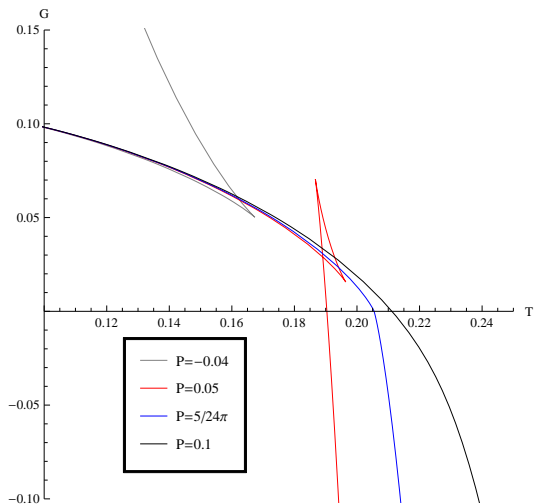}\label{G4}}
\caption{The Gibbs free energy of Gauss-Bonnet black hole as functions  of  $r_{+}$ for different $P$. We have set $\alpha=0.05$.\label{Ggb}
}
\end{figure}
We can also derive the Gibbs free energy for Gauss-Bonnet black holes,
\be
G=-\frac{\pi  \left(-72 \alpha ^2+3 r_+^4 (48 \pi  \alpha  P-1)+4 \pi  P r_+^6+18 \alpha  r_+^2\right)}{24 \left(4 \alpha +r_+^2\right)}.
\ee

From Fig.\ref{G3}, one can see that for constant $r_{+}$, the larger the pressure is, the more stable the Gauss-Bonnet black hole is. While for constant pressure, the larger black hole
is more stable. The Gibbs free energy as a function of temperature for various pressures is shown in Fig.\ref{G4}. Clearly, when $P<P_c=5/24\pi$ the Gibbs free energy with respect to temperature develops a ``swallow tail". At the critical pressure $P=P_c$, the ``swallow tail" disappears which corresponds to the critical point. When $P<0$, the $G-T$ plot is similar to that of RN black hole and that of black hole in de Sitter space\cite{Kubiznak:2015}.

Eq.(\ref{GB11}) can also give an equation of state,
\be
P=\frac{3 \left(44\times 2^{3/4} \pi ^{3/2} \alpha  T+22 \sqrt[4]{2} \sqrt{\pi } T \sqrt{V}-\sqrt[4]{V}\right)}{8 \sqrt{2} V^{3/4}}
\ee
or $P=P(r_{+},T)$.

\begin{figure}[!htbp]
\includegraphics[width=7cm,keepaspectratio]{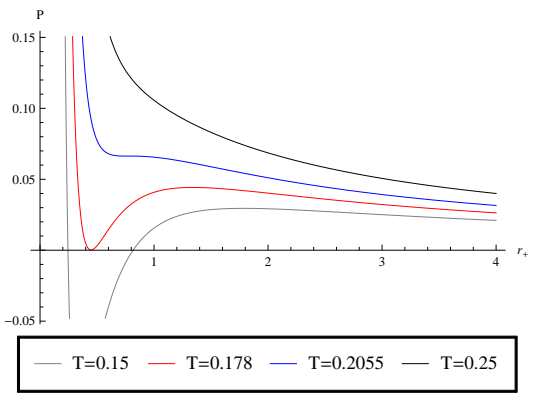}
\caption{The pressure $P$ of Gauss-Bonnet black hole as functions  of  $r_{+}$ for different $T$. We have set $\alpha=0.05$. And $T=0.2055$ is the critical temperature.\label{prgb}
}
\end{figure}

Based on this equation, one can  analyze the $P-V/r_{+}$ criticality of Gauss-Bonnet black hole. In fact, the result (shown in Fig.\ref{prgb}) is similar to that in \cite{Cai:2013} with $Q=0,~k=1,~d=5$.
The $P-V/r_{+}$ criticality is similar to that in Van der Waals liquid/gas system with the critical point at $P_c=\frac{1}{96 \pi  \alpha },~T_c=\frac{1}{4 \sqrt{3} \pi  \sqrt{\alpha }},~r_c=2 \sqrt{3} \sqrt{\alpha }$. In the $P-V$ curves below the critical temperature there exist a portion where $\d{\p P}{\p V}>0$ and should be replaced by an isobar obtained according to Maxwell's equal-area law.

\section{Concluding remarks}

In the framework of horizon thermodynamics, we studied the thermodynamic stabilities of the black holes in GR and GB black hole. In static, spherically symmetric case, whatever the matter fields are, one can always set $P=T^1_{~1}$ as the  pressure of black hole thermodynamic system to obtain the thermodynamic identity $dE=TdS-PdV$ from the field equations. We study the thermodynamic stabilities of black holes according to the variables $E,~T,~S,~P,~V$ by taking the same method and criterion as those in the usual thermodynamics. Therefore, the concrete matter fields are not important. Our discussion applies for all the black holes in the gravitational theories under consideration.

It is shown that in GR only when the radial pressure $P$ is positive, the black holes are stable, because only in this case there exist $\left.\d{\p P}{\p V}\right|_{T}\leq 0$ and $C_P \geq 0$.  In the usual thermodynamic system, $P>0$ is a natural requirement, however in gravitational system it is not the case, such as negative pressure for dark energy.  Thus, we can find that nearly all the static, spherically symmetric solutions in GR, such as RN black hole, Born-Infeld black hole, etc. are thermodynamically unstable.  This result is very different from that obtained in the usual black hole thermodynamics based on the first law $dM=TdS+\phi dQ+...$\cite{Pavon,Chamblin,Lemos,Wu1,Lundgren,Banerjee,Wu2}. This difference can be easily understood. Because the two ways studying the thermodynamic properties of black holes are based on different thermodynamic variables. However, theoretically we cannot conclude which choice is more reasonable.

The thermodynamic stability of GB black hole is also relevant to the pressure $P$. In particular, when $P<0$,  GB black hole can also be thermodynamically stable . We also studied the critical behaviors of GB black hole employing the Gibbs free energy $G$ and the equation of state $P=P(V,T)$. We find that the critical behaviors of GB black hole in the framework of horizon thermodynamics are similar to those discussed in \cite{Cai:2013,LYX}, where the usual black hole thermodynamics is used and the pressure is taken to be $P=-\Lambda/8\pi$.

\bigskip

\section*{Acknowledgements}
MSM would thank Prof. Padmanabhan for useful correspondences. This work is supported in part by the National Natural Science Foundation of China under Grants
 Nos.(11475108) and by the Doctoral Sustentation Fund of Shanxi Datong
University (2011-B-03).

\end{document}